# A novel ZnO piezoelectric microcantilever energy scavenger: Fabrication and characterization

Deepak Bhatia[1,3], Himanshu Sharma[2], R.S.Meena[3], V.R Palkar[1]

[1]Department of Electrical Engineering and Centre for Research in Nanotechnology and Science, Indian Institute of Technology Bombay, Mumbai-400076, India.
[2]Department of Physics, Indian Institute of Technology Bombay, Mumbai-400076, India
[3]Department of Electronics Engineering, Rajasthan Technical University, Kota-324010, India.

**Abstract**

This novel piezoelectric zinc oxide (ZnO) thin film microcantilever was fabricated by micromachining technique. To release the cantilever, wet anisotropic etching of Silicon (Si) was performed by tetramethyl ammonium hydroxide (TMAH). The transverse piezoelectric coefficient $d_{31}$ of the ZnO film, obtained from the deflection of the cantilever with influence of applied voltage, was calculated as 3.32 pC/N. The observed dynamic characterization of the novel piezoelectric microcantilever had linear response with the applied driving voltage. The obtained values of Young Modulus and Hardness were 208±4 GPa and 4.84± 0.1 GPa respectively. This inexpensive novel method provides additional design flexibility to fabricate vibrational energy harvesters. The easy steps of fabrication and cost effectiveness of this method may be preferred it over DRIE. The voltage induced due to deformation of ZnO cantilever were measured ~230mV. This microcantilever energy scavenger may be used to power the nano devices and sensors for medical and agricultural applications as a replacement of traditional bulky batteries.

**Keywords:** ZnO, Tetramethyl ammonium hydroxide (TMAH), Energy Scavengers, Piezoelectric,

## 1. Introduction

With advancement in technology, the power demand of individual devices has drastically come down. Therefore energy scavenging by converting vibration energy into useful output electrical power is looked upon as promising solution. Mechanical vibrations are more popular compared to other available ubiquitous ambient energy sources. There availability is almost everywhere in environment and easily converted in usable electrical power. There have been a lot of efforts in the past for harvesting the low frequency vibrations with micromachined piezoelectric (PZT) cantilevers using integrated proof mass. Reshaping the geometry of the rectangular cantilever to triangular or trapezoidal for increasing average value of strain have also been experimented [1-2]. These methods of tuning the frequencies have lot of process limitations and their fabrication steps are also challenging. Jinhui Song et.al reported power generation process from single ZnO Belt/Wire [3]. Seok-Min Jung et.al proposes an energy-harvester based on the principle of mechanical frequency-up conversion by snap-through

Email: dbhatia@rtu.ac.in

buckling [4]. The research scenario in the area of nano mechanical energy conversion was greatly influenced by Z.L Wang et.al. They have done pioneering work by introducing ZnO nanowires for producing energy at nano scale and a series of devices based on ZnO nanowires such as DC power nanowire generators (NWG), vertically integrated NWG, laterally packed NWG and self powered nano devices were introduced [5–9]. However those nanowires based nano power generators proven their efficiency in scavenging energy from ambient sources, but they suffers with output instability. Other major challenges were reported as mechanical robustness, adaptability and their lifetime, alignment of individual nano devices. Cantilever structure is most suitable for energy scavenging purpose due to its lowest stiffness for a particular size and hence very easy to design a system for low frequencies. That may produce highest average strain to a particular load. This paper focuses on the fabrication and characterization of novel ZnO thin film microcantilever (500 × 100 × .3 $\mu m^3$) energy scavenger using the simple and inexpensive wet etching method. To convert electrical power into mechanical energy, i.e. force and displacement, the converse piezoelectric effect has been utilized in piezoelectric cantilever actuators. Typical applications of piezoelectric cantilever actuators in powering of implantable bio sensors, environmental/agricultural monitoring sensors, ultrasonic motors, explosive detectors, probe tips of atomic-force microscopy (AFM), wireless sensors network and nanodevices with ease and flexibility in operation [10–13].

Piezoelectric materials are perfect candidates for harvesting power from ambient vibration sources. Among the variety of available piezoelectric materials, the most popularly used material is lead zirconate titanate (PZT) due to its superior piezoelectricity [14]. However, poor stability and loss of polarization with continuous usage are the major issues with PZT. Their piezoelectric properties are also strongly affected by operating temperatures and due to brittleness they cannot be deformed mechanically for long duration. Zinc oxide (ZnO) is another important piezoelectric material which is popularly used as one of the pollution-free piezoelectric material and is free from limitations found with PZT. ZnO is highly tensile and may undergo huge mechanical deformations for a long duration without the effect of temperature variation. Therefore it has received increased attention for various Micro Electro Mechanical Systems (MEMS) device applications [15]. Due to the unique combinations of electrical, optical and piezoelectric properties of ZnO, it has great potential for applications in solar cells, photo detectors, and light emitting diodes (LEDs), also it can be easily integrated with other processes and materials. However piezoelectricity of ZnO is generally smaller than that of PZT [16] but has the additional advantage of flexibility in processing. ZnO thin films can be deposited at room temperature and variety of acidic etchants are also available [17]. ZnO is an n-type semiconductor with a wide direct band of 3·3 eV (at room temperature) good electron transporting properties and solution-based processability at low work function. It has a hexagonal quartzite structure and large excitation binding energy of 60 meV which makes ZnO a potential material to realize the next generation MEMS and UV semiconductors [16].

Email: dbhatia@rtu.ac.in

The fabrication and characterization of ZnO thin film cantilever has been reported over past decade in many research papers. They fabricated with the use of Deep Reactive ion etching (DRIE) to release the cantilever. The fabrication of ZnO cantilever with wet etching method is difficult due to sensitivity of ZnO for wet etching and treatment by temperature, acid bases and even water [18-19]. Therefore a novel method was developed for the successful fabrication of ZnO based MEMS devices. The V-grooves in the Si wafer were created by TMAH wet etching before deposition of ZnO layer on substrate [20]. The SEM images of the fully fabricated ZnO cantilever are shown in Fig. 1. A laser Doppler Vibrometer [LDV] Polytec MSA-500 was used to measure the dynamic response of the piezoelectric cantilever. The functioning of the energy scavenger has been tested by Comsol multiphysics software as well as Keithley probe station.

## 2. Modelling and Design

The most popular approach to design the piezoelectric microcantilever in consideration with evenness of moments and forces and the compatibility condition of strain at the piezoelectric and elastic material interface by solving the consecutive equations [21,22]. The actuation force can be calculated with the use of equivalent force and deflection relationship at the tip of the piezoelectric cantilever [23].

Fig. 2 demonstrates a piezoelectric cantilever deflection mechanism by the application of an external load F for a certain displacement δ(l). The speed of positioning has to fulfil the requirements for specified applications in addition to displacement and loading. Thus the most important specifications for the design of piezoelectric cantilever actuators were found to be the force–displacement relationship and resonant frequency [24].

### 2.1 Bending Resonant Frequency

The resonant frequency of the cantilever can be obtained with the solution of eigen value problem of the fourth-order ordinary differential equation in space. The fundamental resonant frequency $f_0$ of a beam which is free at one end and fixed at the other end may be given by [25, 26].

$$f_0 = \frac{.1615}{L^2} \sqrt{\frac{\alpha}{(E_{sub}t_{sub} + E_p t_p)(t_{sub}\rho_{sub} + t_p\rho_p)}} \quad (1)$$

Where, the subscripts 'p' and 'sub' denote the piezoelectric and elastic materials, respectively; and L, t are the length and thickness of the beam, α is given by equation 2

$$\alpha = E_p^2 t_p^4 + E_{sub}^2 t_{sub}^4 + 2E_p t_p E_{sub} t_{sub}(2t_{sub}^2 + 2t_p^2 + 3t_p t_{sub}) \quad (2)$$

Where ρ is density and other parameters are as follows:

Where, L, M, I and E are the length of the beam, mass per unit length area, moment of inertia and Young's modulus, respectively [23]. Per unit length equivalent mass M for the piezoelectric micro cantilever is given by

$$M = W(t_{sub}\rho_{sub} + t_p\rho_p) \quad (3)$$

Where, 'W' is width. Insertion of Equations (3) and 5) and $f_0 = \omega_0/2\pi$ into Eq. (1) gives the resonant frequency $f_0$,

$$\omega_0 = \frac{3.5160}{L^2}\sqrt{\frac{EI}{M}} \quad (4)$$

To calculate the flexural rigidity EI for the microcantilever, which consists of two materials, its cross section can be converted into the corresponding cross section of a single material by the transformed-section method [27]. The equivalent flexural rigidity EI can be calculated by

$$EI = \frac{W\alpha}{12(E_{sub}t_{sub} + E_t t_p)} \quad (5)$$

The transverse piezoelectric strain coefficient $d_{31}$ of a unimorph microcantilever is expressed as [28]

$$d_{31} = -\frac{1}{3} \frac{\alpha \times \partial(l)}{E_{sub} E_p t_{sub}(t_{sub} + t_p) L^2 V} \quad (6)$$

$$d_{31} = -\frac{\partial(l)}{L} \frac{t_p}{V} \quad (7)$$

The tip deflection δ is [16]

$$\delta(l) = \frac{3L^2 t_{sub}(t_p + t_{sub}) E_{sub} E_p V d_{31}}{E_p^2 t_p^4 + E_{sub}^2 t_{sub}^4 + 2E_p t_p E_{sub} t_{sub}(2t_{sub}^2 + 2t_p^2 + 3t_p t_{sub})} \quad (8)$$

The tip deflection δ(l) and resonant frequency $f_0$ are the functions of the piezoelectric cantilever dimensions. Equations (1 & 8) show, that the width w does not affect the resonant frequency and the tip deflection under any external load condition. Under the consideration of external load F, the driving voltage required for a certain tip deflection δ is inversely proportional to the width w. A wider cross section of the piezoelectric cantilever is preferred for lower less power consumption. The effect of length L can be easily seen from equations (1) and (8). The resonant frequency decreases parabolically with L, while the tip deflection increases parabolically with L for F= 0 [29, 30]. The material properties of Si and ZnO [31, 32] are listed in table-1 and the design parameters for the piezoelectric ZnO microcantilever are listed in table-2.

## 3. Experimental Details

The Piezoelectric ZnO microcantilevers were fabricated by micromachining process and patterned by standard optical lithography steps and followed by etching/liftoff of successive layers stack. Tetramethyl ammonium hydroxide (TMAH) is a popular anisotropic silicon (Si) etchant, it contains no alkali metal ions and hence compatible for Micromachining processes [33]. The side walls of the etched Si were defined by the (111) planes. The angle between the (100) plane and sidewalls of etched Si was 54.7º as depicted in Fig. 3. The complete process flow of releasing cantilevers is illustrated in Fig. 4. The Si substrate used was p-type (100) conducting (0.0001-0.0005 Ω cm). The process flow starts with cleaning of silicon wafers by standard Radio Corporation of America (RCA) method. The 1 μm $SiO_2$ layer was grown on Si wafer by wet thermal oxidation method. This layer serves as mask to the Si in the process of TMAH etching of Si [16]. The deposited $SiO_2$ layer is patterned by lithography and BOE etching. For TMAH etch of Si wafer (shallow or deep) 25% TMAH with water was used. The quantity of, TMAH and water in the mixture were 90 ml and 30 ml respectively [18, 19]. At the outset the V-groove was created with the depth of 150 μm on silicon wafer. Then SU-8 2100 of 150 μm thickness was spinned on the wafer and the scraper was used to put the SU-8 uniformly. Now SU-8 covered Silicon wafer was placed on the heater, to remove the bubbles and to smooth the surface by heating. Finally, wafer was dipped in SU-8 developer to the process the SU8 [34, 35].

Further ZnO thin films were deposited by dielectric sputtering (RF magnetron) method using a ZnO target (99.9%) with a 2 inch diameter and 3 mm thickness. During

Email: dbhatia@rtu.ac.in

the deposition of ZnO thin film the RF power was 100 W, the base pressure was $5\times10^{-5}$ mbar and operating pressure was $2.2\times10^{-2}$ mbar. Thin films were deposited in Ar atmosphere with a deposition rate of 5 nm/min. The ZnO cantilever was patterned by standard optical lithographic steps. The final releases of cantilevers were achieved by etching of SU-8 2100 by standard PG remover solution at 70°C.

Therefore complete microfabrication process of the micro cantilever mainly included successive etching of Si and ZnO thin film, top electrode deposition and patterning. The Hysitron, Inc Minneapolis USA make nanoindenter model TI-900, were used for measurement of Young Modulus and Hardness of ZnO thin films.

4. Results and Discussions

The crystalline structures and surface roughness (morphology) of ZnO film were evaluated by X-ray Diffractometer (XRD), scanning electron microscopy (SEM) and atomic force microscopy (AFM), respectively. X-ray from Rigaku (Cu-Kα radiation, λ=1.5405 Å) was used for structural phase identifications. The XRD pattern in Fig.5 indicates that the deposited ZnO film has high diffraction peak located at around $34.422^0$ is very high and which is equivalent to the ZnO (002) peak. So the deposited ZnO thin film on the Si substrate has a c-axis preferred orientation, which is an essential condition for good piezoelectric quality. The depth of the etching of Si was assessed by profilometer (Ambios, USA). The measured depth of the V-groove in Si wafer was 160 μm. To determine the grain and surface morphology of ZnO film, Scanning electron microscopy (SEM) was done using Raith150$^{Two}$ as shown in Fig. 6. It appears from the figure that the grains of the deposited film are uniformly distributed with nearly similar size and very compact. The grain size in the ZnO thin film was found to be of the order of 35-40 nm, with a columnar structure [22, 36]. SEM and profilometer were also used to find the uniformity and the thickness of ZnO films with granular structure. The thickness of the ZnO film was about 0.3 μm.

Further, Fig. 7 shows atomic force microscope (AFM) images of ZnO thin film with 500 nm scan area and 30 nm scanning height. The AFM image showed that the ZnO thin film has very uniform surface, and the roughness of surface was 5.687 nm. The SEM images of the fully fabricated ZnO cantilever are shown in Fig 1, which clearly illustrates that the fabricated device is stable and maintains its freestanding state without structural deformations. The dimension of the ZnO microcantilever is about $500 \times 100$ μm$^2$.

Nanoindentation technique was used to calculate the mechanical properties. Loading unloading of the indenter with standard methods is used to calculate the Hardness and Young modulus of the ZnO films. The hardness of a material indicates its resistance to the elastic deformation. In the nanoindentation process, hardness ($H_{hard}$) is the ratio of maximum indentation load $P_{max}$ to the projected contact area under maximum load ($A_{contact}$), and calculated from:

$$H_{hard} = \frac{P_{max}}{A_{contact}} \qquad (9)$$

Young's modulus $E_s$ of a material is defined in terms of the stiffness or the resistance to plastic deformation and it is calculated from eq:

Email: dbhatia@rtu.ac.in

$$E_s = \left(1 - v_s^2\right)\left(\frac{1}{E} - \frac{1 - v_i^2}{E_i}\right)^{-1} \quad (10)$$

where $v$ is the Poisson's ratio, E refers Young's modulus and the subscripts 'i' and 's' used for indenter and specimen, respectively. The standard parameter values of $E_i$ = 1139 GPa, $v_i$ = 0.071 and vs = 0.401 are used for calculation by assuming indenter probe as a standard diamond probe [37- 39]. The obtained values of Young Modulus and Hardness in GPa units with respect to penetration depth in nm is plotted in the Fig.8. The approximate values of Young modulus of ZnO film is 208±4 GPa and hardness as a function of penetration depth is 4.84± 0.1 GPa. These obtained values of Young's modulus and hardness are almost constant indicates that effect of substrate is not meaningful for these indentation depths.

To evaluate the performance of the fabricated piezoelectric microcantilever actuator, characterization was done by Polytec MSA 500 laser Doppler vibrometer (LDV) [31, 40]. The experimental setup is shown in Fig. 9 and 10. The dynamic response of the fabricated ZnO microcantilever was captured using LDV, and then the tested data is analyzed and transverse piezoelectric constant $d_{31}$ of the ZnO film was calculated.

RF function generator (Agilent 33120A) was used to supply the driving voltage and deflection of the cantilever's tip is measured by the LDV. Driving voltage and tip deflection signals are processed by a dual channel FFT spectrum analyzer through data analysis and evaluation signal processing software Intuitive 8.8 (PSV 8.8). The package of PSV data acquisition software has an analyzer of high featured in time domain. For a variety of input wave forms it provides Zoom Fast Fourier transform (FFT), averaging and peak hold measurements. Its high resolution data visualization in three dimensions (3D) includes full frequency response (FRF), deflection shape (ODS) and function of operational capabilities. At the input, sinusoidal waveforms of different driving voltages are applied and corresponding time responses were investigated. A wide band amplifier 7802 M also integrated with system for proper amplification of the detected signals. The resonant frequency of the microcantilever actuator was measured by an impedance analyzer (Agilent 4294A) and transverse piezoelectric coefficient, actuation sensitivity, bandwidth and nonlinearity were observed.

Fig.11 shows the cantilever's tip velocity magnitude curve as a function of input vibration frequency. It has been observed that the resonant frequency of the microcantilever is 72,312 Hz. When a sine wave signal was applied (AC voltage) by the probe station on the bottom and top electrodes of the ZnO microcantilever, in accordance to piezoelectric effect the cantilever generates vibration. These vibrations of the cantilever were detected by the LDV and the velocity and displacement of the cantilever were measured. The plot of the driving voltage versus measured amplitude of the cantilever tip is given in Fig. 12. In this measurement, the driving voltage frequency was fixed at 10 kHz. From Fig. 12, it was observe that when the driving voltage increases from 1 to 15 V, the amplitude of the ZnO cantilever tip increased from 5.63 to 84.2 nm. It has been observed that amplitude of the cantilever tip increases linearly with the driving voltage.

For a bimorph piezoelectric cantilever structure included one piezoelectric layer and one elastic layer, if both of them

Email: dbhatia@rtu.ac.in

have same width, the tip deflection δ(l) of the cantilever may be expressed a equation 8 [41, 42].

Where, L is the length of the cantilever i.e. distance from fixed end; δ(l) is the deflection; $t_p$ and $t_{sub}$ are the thicknesses of both piezoelectric and elastic layers (Si substrate), respectively and $E_p$ and $E_{sub}$ are the Young's modulus of their materials, respectively; V is the applied voltage; and $d_{31}$ is the transverse piezoelectric constant of the piezoelectric material. In this device, since the ZnO microcantilevers were fabricated directly on conducting Si, and by TMAH etching out of the Si from the substrate, therefore only the thickness of ZnO film has been considered. Owing to the large volume of the material it is easy to measure the mechanical properties of the Si substrate and the properties are also stable. While in the case of thin films, estimation of the properties is difficult due to small volume. Using the criteria of linear dependency of the deflection on the applied voltage, the piezoelectric coefficient $d_{31}$ of the ZnO can be calculated by the use of equation 7 [43 -46].

All the parameters except $d_{31}$ are known for our fabricated ZnO microcantilever, L is the length of the cantilever which is 500 μm, $t_p$ is 300nm, $t_{sub}$ is 300 nm, $E_p$ is $2.1 \times 10^{11}$ Pa, and $E_{sub}$ is $1.9 \times 10^{11}$ Pa. δ(l) and V can be obtained from Fig. 12.

The calculated $d_{31}$ of the ZnO film for the different values of driving voltages is shown in figure 12. From Fig. 12, it has been observed that the transverse piezoelectric constant $d_{31}$ of the ZnO film remains constant regarding the variations in different values of driving voltages. The average calculated value of $d_{31}$ is -3.32 pC/N, which is almost in the same order as that of the ZnO bulk material. The value of $d_{31}$ for bulk material is -5.43 pC/N. However the calculated transverse piezoelectric constant $d_{31}$ of deposited ZnO film is little smaller than that of the ZnO bulk material, but it is still high in comparison to other published results [36, 42, 46].

The performance of designed ZnO cantilever was also evaluated with Comsol Multiphysics 5.0 software. In this model ZnO Cantilever was chosen as piezoelectric layer. In this study, the one end of the structure was clamped related to mechanical boundary conditions, hence fixed constraints conditions were applied on the vertical faces. The vertical faces include piezo (ZnO) and non piezo (Au/Cr) layers. Remaining faces were free and allowed to bend the beam with the application of force or voltage. The deflection of the ZnO cantilever beam under the influence of force is shown Fig. 13 on color coding scheme.

To verify the compression (upward movement) and extension (down) of ZnO cantilevers and to measure the produced voltage, following experiment was done. Two micromanipulators of Keithley probe station was used to connect the top and bottom electrode for voltage measurement. Another one micromanipulator was used to deflect the ZnO cantilever. The free end of cantilever was bent by the moving of micro screw one division. The change in voltage from deflection of cantilever was observed and measured shown in the Fig 14.

Further, it will also intrusting to investigate the piezoelectric properties of the microcantilever harvester, fabricated using the ZnO film integrated with another multifunctional multiferroic thin films in order to enhance the properties and storage of generated voltage. The further research is also in the direction to tune the cantilever frequency as low as environmental vibrations (<50 Hz) to harvest low frequency. This can be achieved easily by change in dimensions and design of cantilever beam.

## 5. Conclusion

This research paper deals with the fabrication and characterization of ZnO piezoelectric microcantilevers by the simple and inexpensive wet etching method. Most importantly, it was demonstrated a simple cost effective platform for development of Vibration based energy scavengers. Multiferroic/ZnO composite microcantilever energy harvesters are also under investigation with a change in dimensions and design to increase and store the generated energy from vibrations.

The obtained values of Young Modulus and Hardness are 208±4 GPa and 4.84± 0.1 GPa respectively.

The transverse piezoelectric constant $d_{31}$ of the ZnO film calculated data obtained from LDV is -3.32 pC/N which is comparatively higher than other reported results [10, 11, 23]. With optimization of the deposition conditions of ZnO thin film (i.e. RF power, $O_2/(Ar+O_2)$ gas ratio) and modification in fabrication process, the higher values of transverse piezoelectric constant $d_{31}$ of the ZnO film may be obtained. This ZnO thin film microcantilever energy scavenger route opens the possibility of innovative research and may be the future replacement for traditional bulky batteries in sensing applications.

## Acknowledgement

The authors wish to acknowledge the partial funding (Grant No. 08DIT006 and 13DIT006) received from the Department of Information Technology, Government of India, through the Centre of Excellence in Nanoelectronics, IIT Bombay and CENSE IISC, Banglore. The authors also acknowledge to Department of Physics and Department of Material Science, IIT Bombay and Rajasthan Technical University, Kota for experimental help.

Email: dbhatia@rtu.ac.in

Email: dbhatia@rtu.ac.in

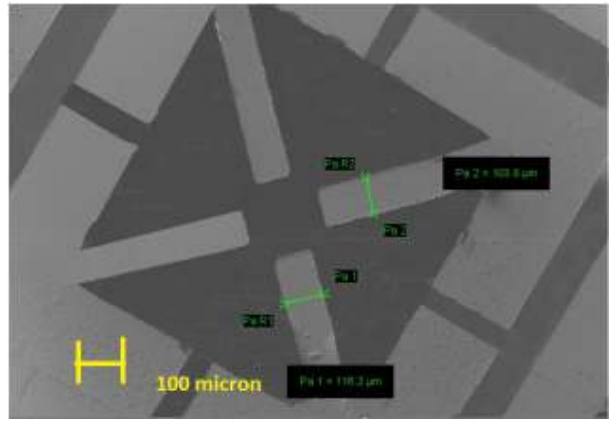

**Figure 1.** SEM image of fabricated ZnO microcantilever.

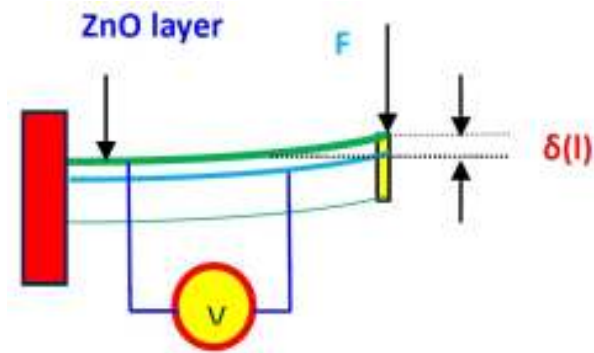

**Figure 2.** Cantilever deflection mechanism

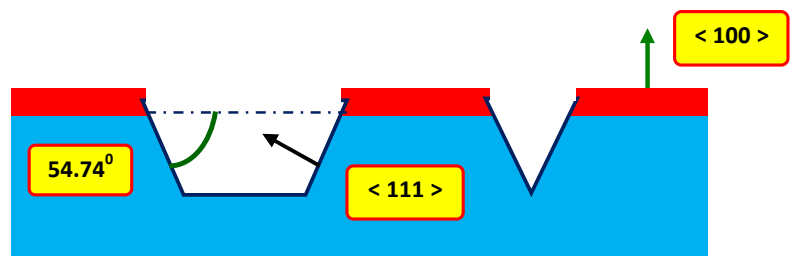

**Figure 3.** Side Profile of TMAH etching of Si (Anisotropic etching)

Color Scheme for Figure 4

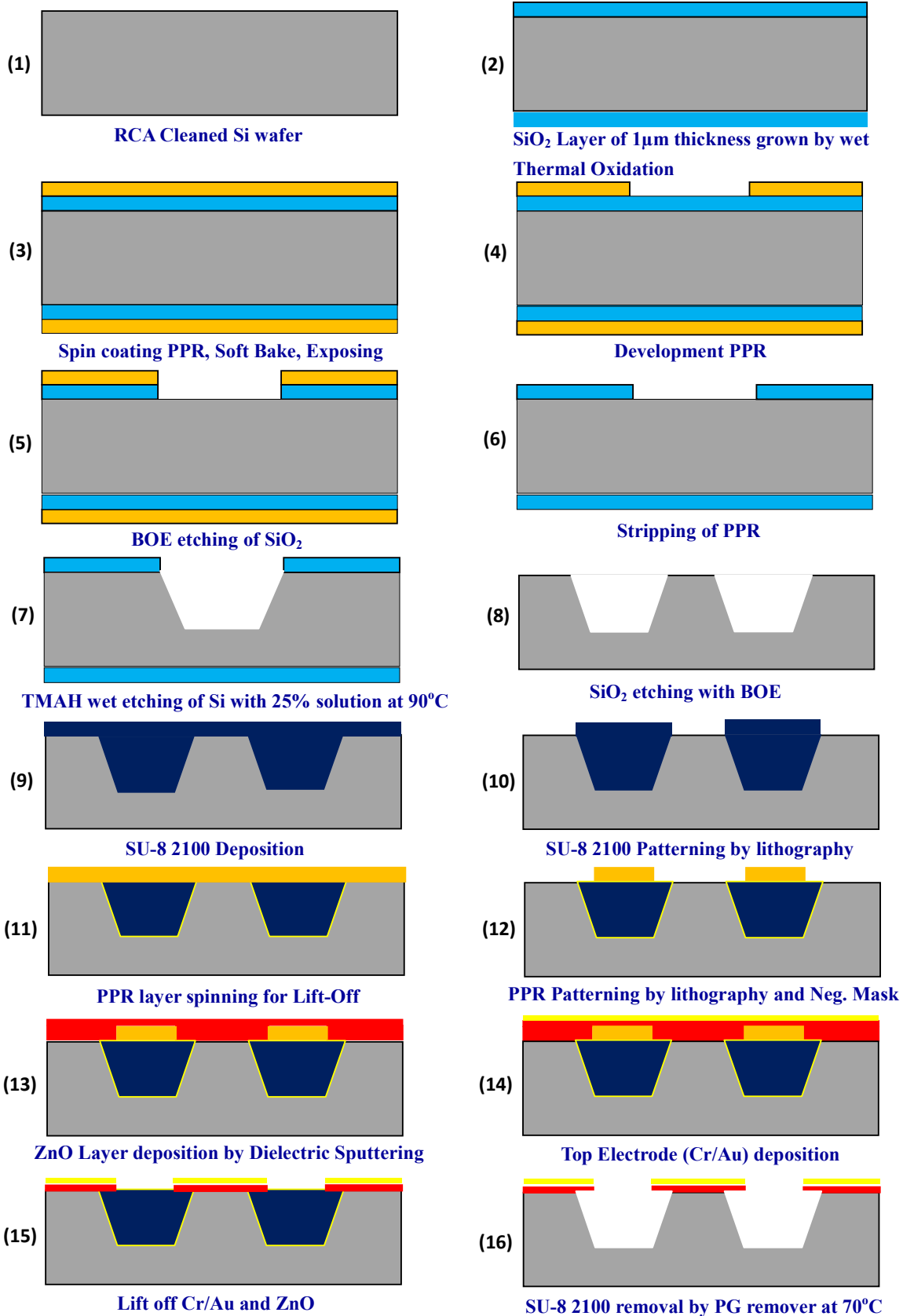

**Figure 4.** Fabrication steps used in the fabrication of Piezoelectric ZnO microcantilevers.

Email: dbhatia@rtu.ac.in

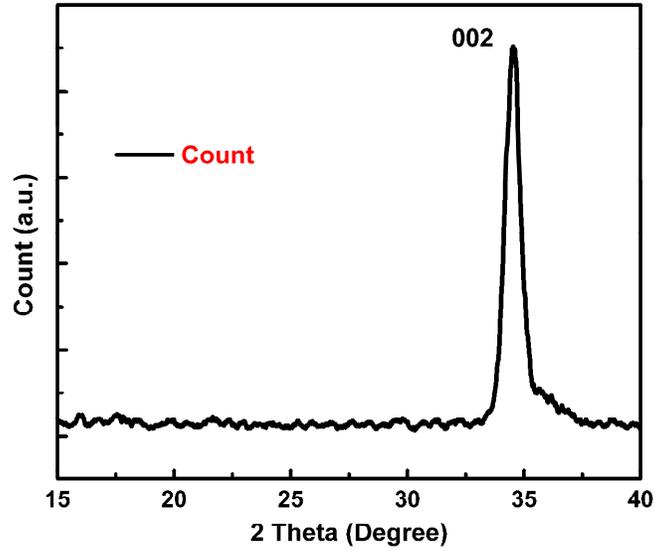

**Figure 5.** X-ray diffraction pattern obtained of ZnO film grown on Si substrate

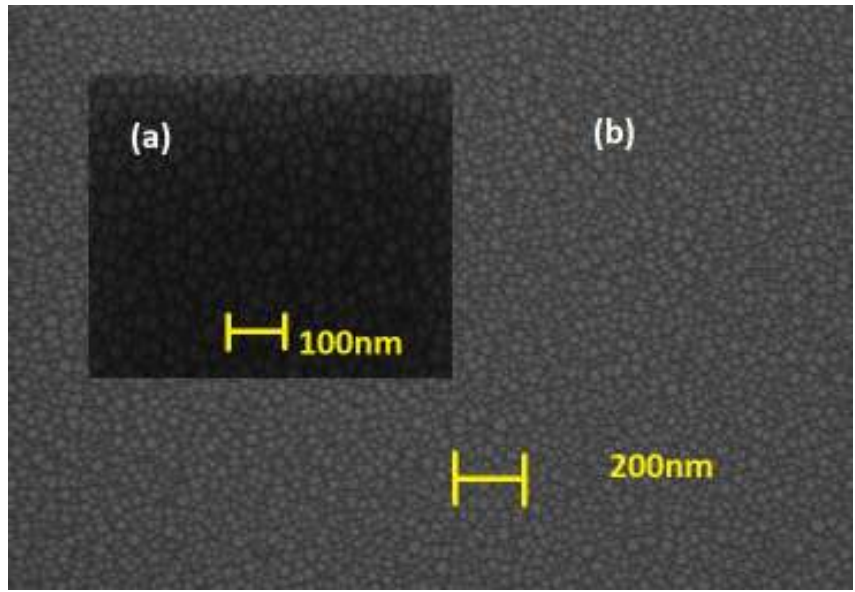

**Figure 6.** SEM image of ZnO Layer at different magnification (a) Inset shows the image at 100 nm (b) at 200 nm

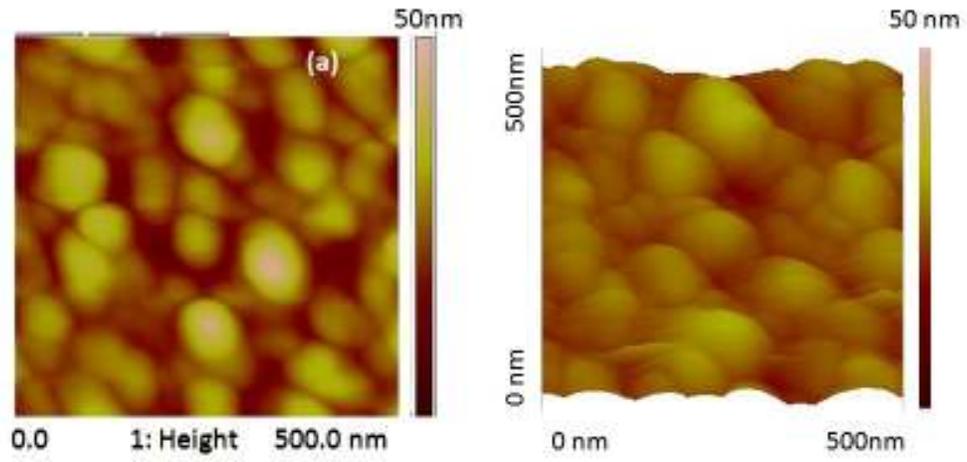

**Figure 7.** AFM (Topography) Images of ZnO thin films of 300nm for scan height30 nm (a) 2 D view(b) 3D View.

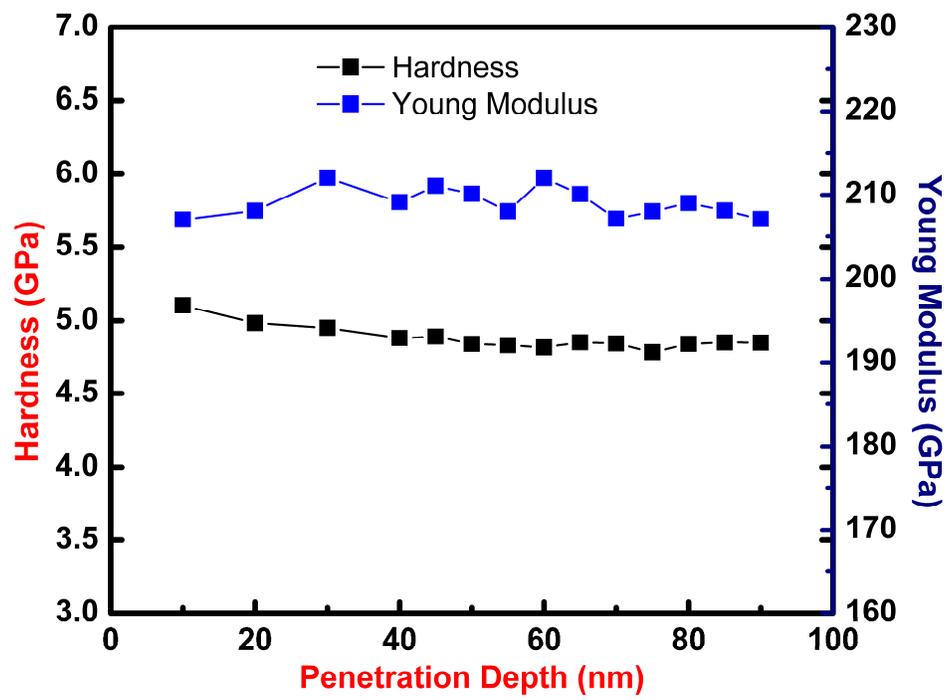

**Figure 8.** Hardness and Young Modulus of ZnO film with Penetration Depth

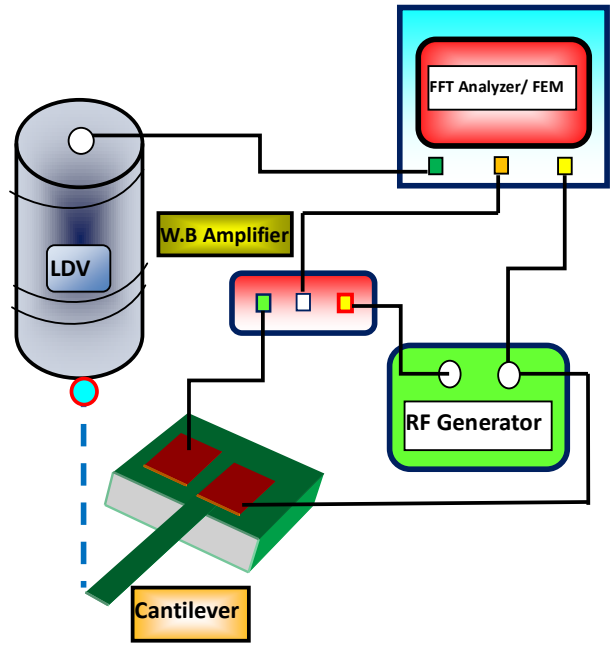

**Figure 9.** LDV (Laser Doppler Vibratometer) setup used for the characterization of piezoelectric ZnO-microcantilever.

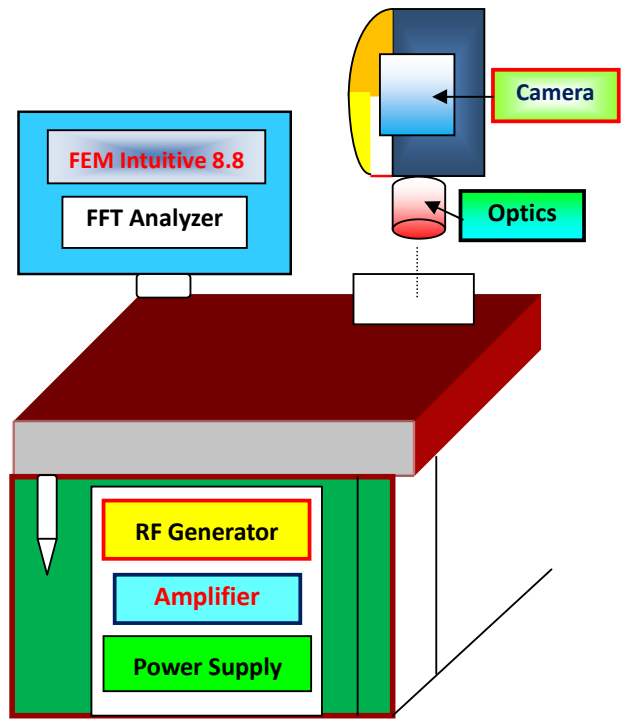

**Figure 10.** LDV (Laser Doppler Vibratometer) FEM and FFT setup for characterization of the piezoelectric ZnO microcantilever.

Email: dbhatia@rtu.ac.in

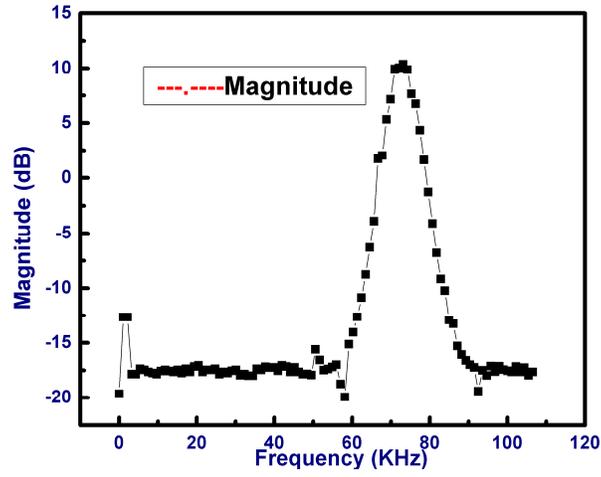

**Figure 11.** The cantilever's tip velocity magnitude variations with input vibration frequency.

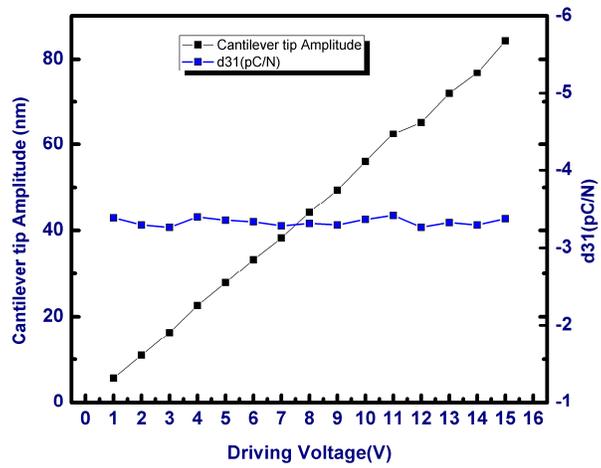

**Figure 12.** Cantiliver tip's measured amplitude and $d_{31}$ of ZnO thin film v/s Voltage applied on microcantilever (measurement was taken on driving AC voltage frequency 10 kHz).

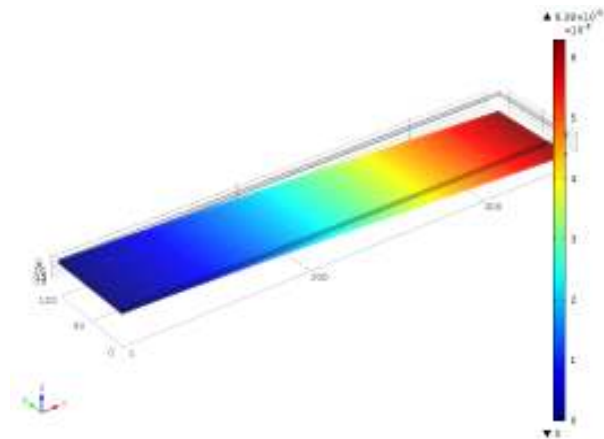

**Figure 13.** Bending of ZnO Cantilever beam under application of force, Electromechanical Characterization

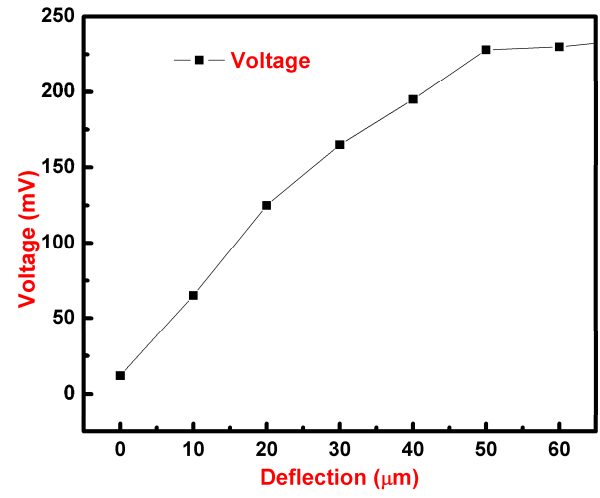

**Figure 14.** Voltage induced by flexing up and down of the cantilevers

**List of figures**

Table- 1
The basic properties of materials Zinc Oxide and Silicon

| Name of Material | E (GPa) | ρ (kg/m³) | Hardness (GPa) | $d_{31}$ (C/N) |
|---|---|---|---|---|
| Zinc Oxide (ZnO) | 208 | 5740 | 4.90 | $-5.43 \times 10^{-12}$ |
| Silicon (Si) | 189 | 2320 | 8.68 | NA |

Table -2
Design Parameters for the piezoelectric ZnO microcantilever.

| Parameters | Calculated Value |
|---|---|
| Resonant frequency ($f_0$) | 70 (kHz) |
| Sensitivity | 4.8 (nm/V) |
| Thickness of ZnO | 300 (nm) |
| Thickness of top Electrode | 20/80 (nm) Cr/Au |
| Length (L) of cantilever | 500 (μm) |
| Width (W) of cantilever | 100 (μm) |

Email: dbhatia@rtu.ac.in